\documentclass[aps,pre,reprint,groupedaddress]{revtex4-2}

\usepackage{amsmath,amssymb}
\usepackage{graphicx}
\usepackage{hyperref}
\usepackage{amsthm}
\newtheorem{theorem}{Theorem}[]
\newtheorem{lemma}[theorem]{Lemma}

\usepackage{verbatim}
\usepackage[dvipsnames]{xcolor}
\DeclareMathOperator{\diag}{diag}
\DeclareMathOperator{\id}{id}

\DeclareMathOperator{\sgn}{sgn}

\begin{document}


\title{Generating functions for message-passing on weighted networks: directed bond percolation and SIR epidemics}


\author{Christoph Widder}
\affiliation{Albert-Ludwigs-Universit\"at, 79104 Freiburg im Breisgau, Germany}
\author{Tanja Schilling}
\affiliation{Albert-Ludwigs-Universit\"at, 79104 Freiburg im Breisgau, Germany}

\date{\today}

\begin{abstract}
We study the SIR (``susceptible, infected, removed/recovered'') model on directed graphs with heterogeneous transmission probabilities within the message-passing approximation. We characterize the percolation transition, predict cluster size distributions and suggest vaccination strategies. All predictions are compared to numerical simulations on real networks. The percolation threshold which we predict is a rigorous lower bound to the threshold on real networks.  For large, locally tree-like networks, our predictions agree very well with the numerical data.
\end{abstract}


\maketitle


\section{Introduction}
To model the spread of a disease across a population, in principle, one needs to solve the corresponding master equation. However, this is possible only for populations with a very simple connectivity pattern, which is hardly ever given in nature. Therefore researchers usually resort to compartment models, such as e.g.~the SIR (``susceptible, infected, removed/recovered'') model and its numerous variations \cite{bailey1975,New2002_2,New2003_2,Vaz2003,Ken2007,Dor2008,Kar2010_2,
Kar2014,pastor2015,newman2018}.

The population of humans across the globe forms a social network in which 
individuals are connected locally in highly correlated clusters, which are then connected to each other in higher layers. This complex structure needs to be taken into account when one analyzes an epidemic model. Combinations of methods from the theory of random graphs and of epidemic modeling have therefore gained in popularity over past 20 years\cite{New2002_2,newman2018,pastor2015}.

One aspect of infectious diseases, which is of particular interest, is the probability of encountering an outbreak across the entire population. In terms of statistical physics, such outbreaks are percolation events, i.e.~events in which an infinitely large sub-network forms across which the disease is ``transported'' (in analogy to the transport of masses or charges across physical networks). Percolation has been a topic of research in statistical physics and mathematics for about 50 years \cite{bollobas2006}. However, there the interest lay to a large extend on percolation on 
lattices or in continuous space and, in particular, on universal critical 
properties rather than on networks. In the context of epidemic modeling, 
percolation needs to be studied on structured and directed networks.

In recent years much progress has been made by modeling effects of non-trivial properties such as degree correlations \cite{New2002,Vaz2003,Bog2005,Gol2008}, clustering \cite{Mil2009,New2009,Gle2010,Kar2010,
Cui2019,Man2020,Man2020_3} and multiplexity \cite{Azi2014,Dom2016,Hac2016,Cel2016,Man2020,Man2020_3} on percolation. In some cases analytical solutions can be obtained while more complex networks are often treated as locally tree-like in order to derive estimates and bounds. In particular, the formalism of generating functions \cite{New2001,New2002,New2002_2,Mey2006,Ken2007,New2009,Kar2010,
Wid2019,Man2020,Man2020_3} and the message-passing technique \cite{Kar2010_2,Cel2016,Tim2017,All2019,Can2019,Min2020} are powerful tools to tackle percolation problems on random and real networks. Further, numerical simulations are used to explore critical phenomena on complex networks \cite{Cai2015,Dom2016,Cui2019}, and beyond the SIR model, some generalized contagion processes \cite{Jan2004,Dod2004} as well as the spread of multiple pathogens \cite{Cai2015,Cui2019,Man2020,Min2020} have been investigated.

Related research examines the significant effect of edge-weights on disease spreading \cite{pastor2015,Sch2007,Yan2008,Chu2011,Bri2011,Yan2012,
		Rat2013,Kam2013,Sun2014,
                Wu2016_2,Sun2017,Spr2019,Bax2021} mainly via mean field and pairwise approximations on configuration model networks which in particular lead to highly efficient immunization strategies \cite{Pen2010,Dei2011,Wan2014}.

            Here, we present an analysis of percolation on directed graphs with heterogeneous occupation probabilities and its application to the late-time behavior of SIR epidemics by means of the message passing approach. Our work generalizes the work by Karrer and co-workers for undirected networks \cite{Kar2014} as well as the work by  Tim\'ar and co-workers for directed networks with equal occupation probabilities \cite{Tim2017}.

\section{Generating functions}\label{sec:generating_functions}
As we will follow the strategy introduced by Newman and co-workers\cite{New2001,New2002_2} and use generating functions to tackle the percolation problem on random graphs, we briefly recall some properties of generating 
functions.
Let $\mathbf{a}\in\mathbb{N}_0^N$ be a random variable with distribution $p(\mathbf{a})$. The probability generating function (PGF) $F:\;\mathbb{R}^N \rightarrow \mathbb{R}$ is defined by
\begin{equation}\label{equ:def_pgf}
	F(\mathbf{x}) := \sum_{\mathbf{a}\geq\mathbf{0}} p(\mathbf{a})\cdot\mathbf{x}^\mathbf{a} \, ,
\end{equation}
where multi-index notation is used. This definition naturally includes PGFs for joint distributions, e.g. the PGF $F:\; \mathbb{R}^N\times\mathbb{R}^M\rightarrow \mathbb{R}$ for the distribution of two random variables $\mathbf{a}\in\mathbb{N}_0^N, \mathbf{b}\in\mathbb{N}_0^M$ is defined by
\begin{equation*}
	F(\mathbf{x}, \mathbf{y}) :=  \sum_{(\mathbf{a},\mathbf{b})\geq\mathbf{0}} p(\mathbf{a},\mathbf{b})\cdot\mathbf{x}^\mathbf{a}\cdot\mathbf{y}^\mathbf{b} \, .
\end{equation*}
The PGFs for the random variables $\mathbf{a}$ and $\mathbf{b}$ are given 
by $F(\mathbf{x},\mathbf{1})$ and $F(\mathbf{1},\mathbf{y})$, respectively. If $N=M$, the PGF for the sum $\mathbf{c}=\mathbf{a}+\mathbf{b}$ is given by $F(\mathbf{x},\mathbf{x})$.

In order to derive the message-passing equations, two properties of PGFs are necessary. 

First, let $\mathbf{a},\mathbf{b}\in\mathbb{N}_0^N$ be \textit{independent} random variables with PGFs $F,G$. The PGF $H$ for the sum $\mathbf{c}:=\mathbf{a}+\mathbf{b}$ is given by
\begin{align*}
	H(\mathbf{x}) &:= \sum_\mathbf{c}p(\mathbf{c})\cdot\mathbf{x}^\mathbf{c}\\
	&= \sum_\mathbf{c} \sum_{\mathbf{a},\mathbf{b}} \delta(\mathbf{c}-\mathbf{a}-\mathbf{b}) p(\mathbf{a})p(\mathbf{b})\cdot \mathbf{x}^\mathbf{a}\mathbf{x}^\mathbf{b} \\
	&=F(\mathbf{x})\cdot G(\mathbf{x}) \, .
\addtocounter{equation}{1}\tag{\theequation}\label{equ:pgf_property1}
\end{align*}

Secondly, consider the following random experiment. Draw a random sequence $\mathbf{a}\in\mathbb{N}_0^N$ from the distribution $p(\mathbf{a})$ with PGF $F$. Then, for each $i=1,\dots,N$ draw a random variable $\mathbf{b}\in\mathbb{N}_0^M$ from the distribution $p_i(\mathbf{b})$ with PGF $G_i$. All random variables be independent. Now, equ.~(\ref{equ:pgf_property1}) yields the PGF $H$ for the random variable $\mathbf{c}:=(\mathbf{a},\mathbf{B})$, where $\mathbf{B}:=\sum_{k=1}^{|\mathbf{a}|}\mathbf{b}_k$:
\begin{align*}
		H(\mathbf{x},\mathbf{y}) &:= \sum_{(\mathbf{a},\mathbf{B})} p(\mathbf{a},\mathbf{B})\cdot \mathbf{x}^\mathbf{a}\cdot\mathbf{y}^\mathbf{B} \\
		&= \sum_\mathbf{a} p(\mathbf{a})\cdot\mathbf{x}^\mathbf{a} \sum_\mathbf{B} p(\mathbf{B})\cdot \mathbf{y}^\mathbf{B} \\
		&=\sum_\mathbf{a} p(\mathbf{a})\cdot\mathbf{x}^\mathbf{a}\prod^N_{i=1} \left[ G_i(\mathbf{y})\right]^{a_i}\\
		&= F(\mathbf{x} * \mathbf{G}(\mathbf{y}))  \, ,
\addtocounter{equation}{1}\tag{\theequation}
\label{equ:pgf_property2}
\end{align*}
where $*$ denotes element-wise multiplication.

\section{Message-passing theory}\label{sec:message_passing}
In this section we interpret message-passing approximations as solutions on infinite trees. To ensure a well defined phase transition, we must restrict ourselves to networks with sufficiently many long loops, such that large clusters above the percolation threshold almost surely form infinite clusters on these trees. A detailed discussion on this issue can be found in \cite{All2019}. \\
Let $\mathcal{G}:=\{V,E\}$ be a large directed network, where $V:=\{1,\dots,N\}$ is the set of nodes and $E\subseteq V\times V$ is the set of $M$ directed edges. For each edge $i\rightarrow j \in E$, the edge weight 
equals the occupation probability which is denoted by $p_{i\rightarrow j}$. The goal is to approximate the PGFs for the distribution of finite clusters $\mathbf{a}\in\{0,1\}^N$ for each node in a large network, where $a_j = 1$, if node $j$ is part of the cluster and $a_j = 0$, otherwise. 
The cluster $\mathbf{a}$ represents the set of nodes, which can be reached from the initial node by a path of occupied edges. The PGF for the cluster distribution of node $i$ takes the form
\begin{equation*}\label{equ:pgf_cluster_distribution}
	H_{0i}(\mathbf{x}) = \sum_{\mathbf{a}} p_i(\mathbf{a})\cdot \mathbf{x}^\mathbf{a} \, .
\end{equation*}
The actual approximation of the message-passing approach is to allow for multiple counts of the same node within the cluster. Hence, each finite cluster $\mathbf{a}\in\mathbb{N}_0^N$ is described by the number of occurrences of each node within the cluster, which in terms of spreading processes means that a node can be traversed multiple times regardless of whether the node has been visited in the past. This reduces the complexity significantly and allows for exact solutions, since the distributions for the partial clusters obtained by following each edge become uncorrelated and independent of past events.

In order to formalize the message-passing approximation, let us consider infinite trees $\mathcal{G}(i)$ obtained by recursively following all outgoing edges without returning to the previously visited node. $\mathcal{G}(i)$ contains an infinite number of copies of edges and nodes from the underlying network $\mathcal{G}$. The occupation probabilities be the same as for the corresponding edges in $\mathcal{G}$ and independent for each copy. The PGF for the cluster configurations of node $i$ within the tree $\mathcal{G}(i)$ can be calculated exactly and yields the message-passing approximation for the network $\mathcal{G}$.

Imagine the formation of a cluster of outgoing occupied edges starting from node $i$ within the tree $\mathcal{G}(i)$. First, instead of nodes, we 
count recursively all edges within the cluster by adding up the unit vectors $\mathbf{e}_{i\rightarrow j}\in \{0,1\}^M$ for each occupied edge. For each edge $i\rightarrow j$ which is encountered, the summand $\mathbf{b}\in\mathbb{N}_0^M$ is drawn independently from the distribution
\begin{equation*}
	f_{i\rightarrow j}(\mathbf{b}) = \left\{
	\begin{array}{ll}
	p_{i\rightarrow j} & \mathbf{b}=\mathbf{e}_{i\rightarrow j} \\
	1-p_{i\rightarrow j} & \mathbf{b}=\mathbf{0}  \\
	0 & \text{else}
	\end{array} \right. 
\end{equation*}
with generating function
\begin{equation*}
	(1-p_{i\rightarrow j}) + p_{i\rightarrow j}\cdot y_{i\rightarrow j} \, ,
\end{equation*}
which represents a Bernoulli experiment for the edge occupation. According to equ.~(\ref{equ:pgf_property1}), the PGF for the occupied edges in the first step is 
\begin{equation}\label{equ:def_G0}
	G_{0i}(\mathbf{y};\mathbf{p}) = \prod_{j\in\mathcal{N}^+(i)}(1-p_{i\rightarrow j}) + p_{i\rightarrow j}\cdot y_{i\rightarrow j} \, ,
\end{equation}
where $\mathcal{N}^+(i)$ denotes the set of successors of node $i$. Similar, the PGF for the occupied outgoing edges of node $j$, which do not lead back to node $i$, is given by
\begin{equation}\label{equ:def_G}
	G_{i\rightarrow j}(\mathbf{y};\mathbf{p}) = \prod_{k\in\mathcal{N}^+(j)\setminus i}(1-p_{j\rightarrow k}) + p_{j\rightarrow k}\cdot y_{j\rightarrow k} \, .
\end{equation}
For simplicity, the parameter $\mathbf{p}$ is dropped, where possible. Equ.~(\ref{equ:pgf_property2}) yields the joint PGF for the occupied edges for the first and second step
\begin{equation*}
	G_{0i}(\mathbf{y}_1*\mathbf{G}(\mathbf{y}_2)) \, .
\end{equation*}
By recursively applying equ.~(\ref{equ:pgf_property2}) $n-1$ times, one obtains the PGF for the edges within the cluster for each of the first $n$ 
steps
\begin{equation*}
	G_{0i}\circ [\mathbf{y}_1*\mathbf{G}] \circ \dots \circ [\mathbf{y}_{n-1}*\mathbf{G}] (\mathbf{y}_n) \, .
\end{equation*}
Hence, the PGF for the sum of all edges within the cluster up to the $n$-th nearest neighbors of node $i$ is given by:
\begin{align*}
	& G_{0i}(\mathbf{H}^{(n)}(\mathbf{y})) \\
	& \mathbf{H}^{(n)} = \mathbf{y} * \mathbf{G}(\mathbf{H}^{(n-1)}(\mathbf{y})) \\
	& \mathbf{H}^{(1)} = \mathbf{y} \, .
\end{align*}
This is the analogon to equ.~(46) from ref.~\cite{New2001} for the number 
of the $n$-th nearest neighbors on random graphs.

Now, define $H_{i\rightarrow j}(\mathbf{y}):\,\mathbb{R}^M\rightarrow\mathbb{R}$ to be the PGF for the edges in finite partial clusters when following the edge $i\rightarrow j$, including $i\rightarrow j$. These PGFs are independent for all edges, therefore, according to equ.~(\ref{equ:pgf_property2}) $\mathbf{H}(\mathbf{y})$ satisfies the following fixed-point equation and yields the PGF $H_{0i}(\mathbf{y})$ for all arbitrary large but finite clusters of edges $\mathbf{c}\in\mathbb{N}_0^{M}$  
\begin{align*}
	 H_{0i}(\mathbf{y}) &= G_{0i}(\mathbf{H}(\mathbf{y})) \\
	  \mathbf{H}(\mathbf{y}) &= \mathbf{y} * \mathbf{G}(\mathbf{H}(\mathbf{y})) \, . 
\end{align*}
Finally, we obtain the PGF $H_{0i}(\mathbf{x}):\,\mathbb{R}^N\rightarrow\mathbb{R}$ for the distribution of finite clusters $\mathbf{a}\in\mathbb{N}_0^N$ by multiplying $x_i$ for the root and applying the concatenation $y_{i\rightarrow j}=x_j$ in order to count nodes instead of edges, which  are trivial cases of equs.~(\ref{equ:pgf_property1}) and (\ref{equ:pgf_property2})
\begin{align}
		 H_{0i}(\mathbf{x};\mathbf{p}) &= x_i \cdot G_{0i}(\mathbf{H}(\mathbf{x};\mathbf{p});\mathbf{p}) \label{equ:mp_out_1}\\
	  H_{i\rightarrow j}(\mathbf{x};\mathbf{p}) &= x_j \cdot G_{i\rightarrow j}(\mathbf{H}(\mathbf{x};\mathbf{p});\mathbf{p}) \, . \label{equ:mp_out_2}
\end{align}
These \textit{message-passing} equations fully determine the formation of 
finite clusters, hence, they are sufficient to solve percolation on the tree $\mathcal{G}(i)$. This is completely analogous to equs.~(27) and (26) 
from ref.~\cite{New2001} for the cluster size distribution on random graphs. 

The solutions for percolation of \textit{incoming} occupied edges on the tree $\mathcal{F}(i)$, which is obtained by recursively following all incoming edges without returning to the previously visited node, can simply be obtained by flipping all arrows and introducing new letters without repeating the procedure. With the definitions
\begin{align}
	F_{0i}(\mathbf{y};\mathbf{p}) &= \prod_{j\in\mathcal{N}^-(i)}(1-p_{i\leftarrow j}) + p_{i\leftarrow j}\cdot y_{i\leftarrow j} \label{equ:def_F0}\\
	F_{i\leftarrow j}(\mathbf{y};\mathbf{p}) &= \prod_{k\in\mathcal{N}^-(j)\setminus i}(1-p_{j\leftarrow k}) + p_{j\leftarrow k}\cdot y_{j\leftarrow k} \, , \label{equ:def_F}
\end{align} 
where $\mathcal{N}^-(i)$ is the set of predecessors of node $i$, the PGFs 
for the distributions of finite clusters of incoming edges are given by
\begin{align}
	Q_{0i}(\mathbf{x};\mathbf{p}) &= x_i \cdot F_{0i}(\mathbf{Q}(\mathbf{x};\mathbf{p});\mathbf{p}) \label{equ:mp_in_1}\\
	Q_{i\leftarrow j}(\mathbf{x};\mathbf{p}) &= x_j \cdot F_{i\leftarrow j}(\mathbf{Q}(\mathbf{x};\mathbf{p});\mathbf{p}) \, . \label{equ:mp_in_2}
\end{align}

For $\mathbf{x}=x\cdot\mathbf{1}$ and $p_{i\rightarrow j}\equiv p$, equs.~(\ref{equ:mp_out_1})and (\ref{equ:mp_in_1}) are reduced to equ.~(3) from ref.~\cite{Kar2014} for undirected networks and equs.~(\ref{equ:mp_out_2}) and (\ref{equ:mp_in_2}) are reduced to equs.~(3) and (4) from ref.~\cite{Tim2017} for directed networks by substitution according to 
\begin{align*}
H^{(\text{out})}_{i\rightarrow j}(x):=1-p+p\cdot H_{i\rightarrow j}(x\cdot \mathbf{1}) \\
H^{(\text{in})}_{i\leftarrow j}(x):=1-p+p\cdot Q_{i\leftarrow j}(x\cdot 
\mathbf{1}) \, .
\end{align*} 

\subsection{Percolation probability}\label{sec:percolation_probability}

 Above the percolation threshold, there is a chance, that the cluster will become infinite. The cluster distribution contains the probabilities for all finite clusters, hence, $H_{0i}(\mathbf{1})$ and $Q_{0i}(\mathbf{1})$ are the probabilities that node $i$ is in a finite cluster of outgoing 
and incoming edges, respectively. Thus, according to equs.~(\ref{equ:mp_out_1})(\ref{equ:mp_out_2})(\ref{equ:mp_in_1}) and (\ref{equ:mp_in_2}), the percolation probabilities for a randomly chosen node are given by
\begin{align}
	P_{\text{out}} &= \frac{1}{N}\sum^N_{i=1} P_{\text{out}}(i)  \label{equ:P_out}\\
	P_{\text{out}}(i) &= 1-G_{0i}(\mathbf{H}) \label{equ:P_out_i}\\
	\mathbf{H} &= \mathbf{G}(\mathbf{H}) \label{equ:fp_out}\\
	P_{\text{in}} &= \frac{1}{N}\sum^N_{i=1} P_{\text{in}}(i) \label{equ:P_in}\\
	P_{\text{in}}(i) &= 1-F_{0i}(\mathbf{Q}) \label{equ:P_in_i}\\
	\mathbf{Q} &= \mathbf{F}(\mathbf{Q}) \label{equ:fp_in}\, ,
\end{align}
where $H_{i\rightarrow j}:=H_{i\rightarrow j}(\mathbf{1})$ is the probability that the partial cluster obtained by following the occupied edge $i\rightarrow j$ is finite and $Q_{i\leftarrow j}:=Q_{i\leftarrow j}(\mathbf{1})$ is the probability that the partial cluster obtained by backtracking the occupied edge $i\leftarrow j$ is finite. For $p_{i\rightarrow j}\equiv p$, equs.~(\ref{equ:P_out}) and (\ref{equ:P_in}) are equivalent to equs.~(5) and (6) from ref.~\cite{Tim2017} for directed networks and equ.~(6) from ref.~\cite{Kar2014} for undirected networks. For locally tree-like networks, the probabilities for a node to be part of a giant cluster 
of outgoing and incoming edges are independent, therefore, the probability that a randomly chosen node is part of a giant cluster of outgoing and incoming edges simultaneously is given by
\begin{equation}
	P_S = \frac{1}{N}\sum^N_{i=1} P_{\text{in}}(i)\cdot P_{\text{out}}(i) \, , \label{equ:P_S}
\end{equation}
which - due to the existence of loops - equals the probability that a randomly chosen node is part of the giant strongly connected component within the cluster. For $p_{i\rightarrow j}\equiv p$, equ.~(\ref{equ:P_S}) is equivalent to equ.~(7) from ref.~\cite{Tim2017}.

Let us go one step ahead and consider each copy of node $i$ in $\mathcal{G}(i)$ and $\mathcal{F}(i)$ to be vacant with independent probabilities $q_i\in[0,1)$. The PGFs for the distribution of occupied nodes within finite clusters of occupied edges are
\begin{align*}
	&H_{0i}(\mathbf{q}+(\mathbf{1}-\mathbf{q})*\mathbf{x}) \\
	&Q_{0i}(\mathbf{q}+(\mathbf{1}-\mathbf{q})*\mathbf{x}) \, .
\end{align*}
Hence, the probabilities that node $i$ is part of a \textit{vacant} cluster without any occupied nodes is obtained by inserting $\mathbf{x}=\mathbf{0}$ and therefore given by the PGFs for the distribution of clusters evaluated at $\mathbf{x}=\mathbf{q}$, i.e. $H_{0i}(\mathbf{q})$ and $Q_{0i}(\mathbf{q})$. Clearly, in the limit of large networks with many loops, the probabilities for a node to be part of a vacant cluster obtained by the message-passing approximation must always be smaller than or equal to the \textit{true} probabilities due to the overcount of nodes. 
\begin{align*}
	&H^{\text{true}}_{0i}(\mathbf{q}) \geq H_{0i}(\mathbf{q}) \quad \forall_{\mathbf{q}\in[0,1)^N}\\
	&Q^{\text{true}}_{0i}(\mathbf{q}) \geq Q_{0i}(\mathbf{q}) \quad \forall_{\mathbf{q}\in[0,1)^N}\, .
\end{align*}
In the limit $\|\mathbf{1}-\mathbf{q}\|_\infty \ll 1$, we find
\begin{align}
	P^{\text{true}}_{\text{out}} &\sim \frac{1}{N}\sum^N_{i=1}1-H^{\text{true}}_{0i}(\mathbf{q}) \label{equ:upper_bound_out}\\
	&\leq \frac{1}{N}\sum^N_{i=1}1-H_{0i}(\mathbf{q}) \sim P_{\text{out}} \\
	P^{\text{true}}_{\text{in}}  &\lesssim P_{\text{in}} \, .\label{equ:upper_bound_in}
\end{align}
Thus, the message-passing approach yields a rigorous upper bound for the percolation probabilities and a rigorous lower bound for the percolation threshold for networks with loops in the large $N$ limit. If the overcount of nodes becomes negligible, the message-passing approximation must converge to the exact result in the large $N$ limit, which is the case, if the probability for a finite cluster to contain closed loops vanishes. This is true for locally tree-like networks, except at the percolation 
threshold, where the average finite cluster size diverges and the largest 
finite size effects are expected. 

\subsection{Cluster size distribution}\label{sec:cluster_size_distribution}

The PGFs for the distribution of finite cluster sizes are
\begin{align*}
	H_{0i}(x) &:= \sum_{\mathbf{a}\geq\mathbf{0}} p_i(\mathbf{a})\cdot x^{|\mathbf{a}|} \\
	&= H_{0i}(x\cdot\mathbf{1}) 
\addtocounter{equation}{1}\tag{\theequation}
\label{equ:def_csd_out}\\
Q_{0i}(x) &= Q_{0i}(x\cdot\mathbf{1}) \, .
\addtocounter{equation}{1}\tag{\theequation}
\label{equ:def_csd_in}
\end{align*}
After averaging over all nodes and normalization, we obtain the average size of finite clusters for a randomly chosen node 
\begin{align}
	\langle n_{\text{out}} \rangle &= \frac{\sum^N_{i=1}H_{0i}'(1)}{\sum^N_{i=1}H_{0i}(1)} \label{equ:acs_out_1}\\
	H_{0i}'(1) &= H_{0i}(1) + G_{0i}'(\mathbf{H})\cdot\mathbf{H'} \label{equ:acs_out_2}\\
	\mathbf{H'} &= \mathbf{H}+\mathbf{G}'(\mathbf{H})\cdot\mathbf{H'} \label{equ:acs_out_3}\\
		\langle n_{\text{in}} \rangle &= \frac{\sum^N_{i=1}Q_{0i}'(1)}{\sum^N_{i=1}Q_{0i}(1)} \label{equ:acs_in_1}\\
	Q_{0i}'(1) &= Q_{0i}(1) + F_{0i}'(\mathbf{Q})\cdot\mathbf{Q'} \label{equ:acs_in_2}\\
	\mathbf{Q'} &= \mathbf{Q}+\mathbf{F}'(\mathbf{Q})\cdot\mathbf{Q'}  \, ,\label{equ:acs_in_3}
\end{align}
where $\mathbf{H'}:=\mathbf{H'}(\mathbf{1})\cdot\mathbf{1}$ and $\mathbf{Q'}:=\mathbf{Q'}(\mathbf{1})\cdot\mathbf{1}$. The average finite cluster sizes for node $i$ are given by
\begin{align*}
	\langle n_{\text{out}}(i)\rangle &= H_{0i}'(1)/H_{0i}(1) \\
	\langle n_{\text{in}}(i)\rangle &= Q_{0i}'(1)/Q_{0i}(1) \, .
\end{align*}
These are equivalent to equ.~(8) from ref.~\cite{Kar2014} for undirected networks with $p_{i\rightarrow j}\equiv p$.
Beyond the percolation threshold, we have $\mathbf{H}=\mathbf{Q}=\mathbf{1}$ and $H_{0i}=Q_{0i}=1$, thus
\begin{align}
	\langle n_{\text{out}} \rangle &= \frac{1}{N}\sum^N_{i=1}H_{0i}'(1) \label{equ:acs_out_4}\\
	H_{0i}'(1) &= 1 + G_{0i}'(\mathbf{1})\cdot\mathbf{H'} \label{equ:acs_out_5}\\
	\mathbf{H'} &= \mathbf{1}+\mathbf{G}'(\mathbf{1})\cdot\mathbf{H'} \label{equ:acs_out_6}\\
		\langle n_{\text{in}} \rangle &= \frac{1}{N} \sum^N_{i=1}Q_{0i}'(1) 
\label{equ:acs_in_4}\\
	Q_{0i}'(1) &= 1 + F_{0i}'(\mathbf{1})\cdot\mathbf{Q'} \label{equ:acs_in_5}\\
	\mathbf{Q'} &= \mathbf{1}+\mathbf{F}'(\mathbf{1})\cdot\mathbf{Q'}  \label{equ:acs_in_6}\, .
\end{align}
Within the non-percolating phase, the average cluster sizes obtained by the message-passing approximation must be greater than or equal to the true value, due to the overcount of nodes
\begin{align}
	H_{0i}'(1)^\text{true} &\leq H_{0i}'(1) \label{equ:acs_upper_bound_out}\\
	Q_{0i}'(1)^\text{true} &\leq Q_{0i}'(1) \, .\label{equ:acs_upper_bound_in}
\end{align}
For $\rho(\mathbf{G}'(\mathbf{1}))<1$ and $\rho(\mathbf{F}'(\mathbf{1}))<1$, where $\rho$ denotes the spectral radius, we find
\begin{align*}
	\mathbf{H'} &= (\id - \mathbf{G}'(\mathbf{1}))^{-1} \cdot\mathbf{1} \\
	\mathbf{Q'} &= (\id - \mathbf{F}'(\mathbf{1}))^{-1} \cdot\mathbf{1} \, 
,
\end{align*}
therefore, the average finite cluster sizes possess a singularity at 
$\rho(\mathbf{G}'(\mathbf{1}))=1$ and $\rho(\mathbf{F}'(\mathbf{1}))=1$, respectively. These singularities mark the critical points at which the formation of giant clusters become possible, which is again analogous to the theory of random graphs, see equs.~(31) and (32) from ref.~\cite{New2001} and equ.~(22) from ref.~\cite{New2002_2}.

\subsection{Percolation threshold}\label{sec:percolation_threshold}

First, consider percolation of incoming edges. The percolation threshold is the critical point at which the percolation probability $P_{\text{in}}$ becomes positive. For $\mathbf{p}\in[0,1]^M$, the set of all critical points is defined by
\begin{equation}\label{equ:def_P_c}
	\mathcal{P}_c = \partial\{\mathbf{p} \;|\, P_{\text{in}}=0\}\cap \partial\{\mathbf{p} \;|\, P_{\text{in}}>0\} \, .
\end{equation}
Beyond the percolation threshold, $\mathbf{Q}=\mathbf{1}$ is the trivial solution of the the fixed-point equation $\mathbf{Q}=\mathbf{F}(\mathbf{Q})$. The percolation probability $P_{\text{in}}$ is positive, if and only if at least one component of $\mathbf{Q}$ becomes smaller than one. Thus, for continuous phase transitions, consider the first order expansion of the fixed-point equation for $\mathbf{Q}=\mathbf{1}-\boldsymbol{\epsilon}$
\begin{equation*}
	\boldsymbol{\epsilon} = \mathbf{F}'(\mathbf{1})\cdot \boldsymbol{\epsilon} \, .
\end{equation*}
Following the same line of argumentation as for ordinary percolation \cite{Kar2014}, the trivial solution $\boldsymbol{\epsilon}=\mathbf{0}$ becomes unstable, if the spectral radius of $\mathbf{F}'(\mathbf{1})$ exceeds one, which marks the point at which a non-trivial solution is obtained and 
the percolation threshold is exceeded. Hence, for any $\mathbf{p}\in\mathcal{P}_c$ 
\begin{equation}\label{equ:p_c_in}
	\rho(\mathbf{F}'(\mathbf{1};\mathbf{p})) = 1 \, .
\end{equation}
Similarly, the critical points at which $P_{\text{out}}$ becomes positive 
satisfy
\begin{equation}\label{equ:p_c_out}
	\rho(\mathbf{G}'(\mathbf{1};\mathbf{p})) = 1 \, ,
\end{equation}
which is again analogous to the theory of random graphs, see equ.~(32) from ref.~\cite{New2001}. 
We introduce the Hashimoto-Matrix or non-backtracking matrix \cite{Has1989}
\begin{equation}\label{equ:def_hashimoto}
B^T_{i\leftarrow j, k\leftarrow l}=B_{i\rightarrow j,k\rightarrow l}:=\delta_{jk}(1-\delta_{il}) \, ,
\end{equation}
which is useful in applications such as community detection \cite{New2013,Krz2013} and network centrality  \cite{Mar2014}. 
With this we find
\begin{align}
	\mathbf{F}'(\mathbf{1}) &= B^T\cdot \diag(\mathbf{p}) \label{equ:F_prime}\\
	\mathbf{G}'(\mathbf{1}) &= B\cdot \diag(\mathbf{p}) \, .\label{equ:G_prime}
\end{align}
For $p_{i\rightarrow j}\equiv p$, the well-known percolation threshold $p_c=\rho(B)^{-1}$ is retrieved \cite{Kar2014,Tim2017,Min2020}. 

The percolation thresholds for $P_{\text{in}}$ and $P_{\text{out}}$ are the same, since \begin{equation}\label{equ:rho}
	\rho:=\rho(\mathbf{F}'(\mathbf{1}))=\rho(\mathbf{G}'(\mathbf{1})) \, 
,
\end{equation}
which is proven using the Leibniz formula in lemma \ref{lemma:F_prime_eq_G_prime}. Hence, either equ.~(\ref{equ:p_c_in}) or (\ref{equ:p_c_out}) can be used to derive criteria which prohibit the formation of giant clusters on any large network as illustrated for the SIR model in sec.~(\ref{sec:suppression_large_outbreaks}).

\begin{lemma}\label{lemma:F_prime_eq_G_prime}
	The characteristic polynomials for $\mathbf{F}'(\mathbf{1})$ and $\mathbf{G}'(\mathbf{1})$ are equal.
\end{lemma}

\textit{Proof.} Each permutation $\sigma\in S_M$ can be represented by a concatenation of cyclic permutations $\sigma=\pi_1\circ\dots\circ\pi_n$. Each $\pi_k$ permutes a sequence of distinct indices $I_k=(i_1,\dots,i_{m(k)})$, such that $\pi_k(i_l)=i_{l+1}$, if $i_l\in I_k$ and $\pi_k(i_l)=i_l$, else, where $i_{m(k)+1}:=i_1$. Further, let  $I_0:=\{1,\dots,M\}\setminus\cup^n_{k=1}I_k$. The characteristic polynomial of $\mathbf{G}'(\mathbf{1})$ is given by
	\begin{align*}
	&\sum_{\sigma\in S_M}\sgn(\sigma)\prod^M_{i=1} [B\diag(\mathbf{p})-\lambda\id]_{i,\sigma(i)}\\
		&= \sum_{\sigma\in S_M} (-\lambda)^{|I_0|} \prod^n_{k=1}\sgn(\pi_k)\prod_{i\in I_k} [B\diag(\mathbf{p})]_{i,\pi_k(i)} \\
	\end{align*}
and for $\mathbf{F}'(\mathbf{1})$
	\begin{align*}
	&\sum_{\sigma\in S_M}\sgn(\sigma)\prod^M_{i=1} [B^T\diag(\mathbf{p})-\lambda\id]_{\sigma(i),i}\\
		&= \sum_{\sigma\in S_M} (-\lambda)^{|I_0|} \prod^n_{k=1}\sgn(\pi_k)\prod_{i\in I_k} [B^T\diag(\mathbf{p})]_{\pi_k(i),i} \; .\\
	\end{align*}
However,
\begin{align*}
	\prod_{i\in I_k} [B\diag(\mathbf{p})]_{i,\pi_k(i)} &= \left\{
	\begin{array}{ll}
	\prod_{i\in I_k}p_i & \;, \prod_{i\in I_k}B_{i,\pi_k(i)}=1 \\
	0 & \;, \text{else}\\
\end{array}	 \right. \\
&= \prod_{i\in I_k} [B^T\diag(\mathbf{p})]_{\pi_k(i),i} \; ,
\end{align*}
which concludes the proof. $\square$

In the following, we derive some additional, rigorous results for the percolation threshold for percolation of incoming edges. The same results are obtained for percolation of outgoing edges by replacing $\mathbf{F}$, $\mathbf{Q}$ and $P_{\text{in}}$ with $\mathbf{G}$, $\mathbf{H}$ and $P_{\text{out}}$, respectively. 

\begin{lemma}\label{lemma:rho_geq_1}
	Let $\mathbf{p}(\lambda):[0,1]\rightarrow[0,1]^M$ be a continuous parametrization of the occupation probabilities. If a continuous phase transition occurs at $\lambda_c\in[0,1)$, then
	\begin{equation*}
		\rho(\lambda_c) \geq 1 \, .
	\end{equation*}
\end{lemma}

\textit{Proof.} Since $P_{\text{in}}(\lambda_c)=0$ is continuous in $\lambda_c$, we have $\lim_{\lambda\rightarrow\lambda_c}\mathbf{Q(\lambda)}=\mathbf{1}$, where the components of $\mathbf{Q}$ are defined as the probabilities to obtain a finite partial cluster when backtracking the corresponding edges. $\mathbf{Q}$ solves $\mathbf{Q}=\mathbf{F}(\mathbf{Q})$ and is continuous at $\lambda_c$, thus, we may apply the first order expansion for $\mathbf{Q}=\mathbf{1}-\boldsymbol{\epsilon}$ with $\boldsymbol{\epsilon}\gneq\mathbf{0}$ and w.l.o.g. for the limit from the right $\lim_{\lambda\searrow\lambda_c}\boldsymbol{\epsilon} = \mathbf{0}$
\begin{equation*}
	\|\boldsymbol{\epsilon}\| = \|\mathbf{F}'(\mathbf{1})\cdot \boldsymbol{\epsilon}\|+o(\|\boldsymbol{\epsilon}\|) \, .
\end{equation*}
For any induced matrix norm we have
\begin{equation*}
	\|\mathbf{F}'(\mathbf{1})\cdot \boldsymbol{\epsilon}\| \leq \|\mathbf{F}'(\mathbf{1}) \| \cdot \| \boldsymbol{\epsilon}\| \, ,
\end{equation*}
hence, for $\lambda\searrow\lambda_c$, we find
\begin{equation*}
	 \|\mathbf{F}'(\mathbf{1};\lambda_c)\| \geq 1 \, .
\end{equation*}
Further, for any $\varepsilon>0$ exists an induced matrix norm, such that
\begin{equation*}
	\rho(\lambda_c)+\varepsilon \geq \|\mathbf{F}'(\mathbf{1};\lambda_c)\| \geq 1 \, ,
\end{equation*}
which yields a contradiction for $\rho(\lambda_c)<1$. $\square$

For irreducible Hashimoto matrices $B$, the expression for the percolation threshold is a consequence of the Perron-Frobenius theorem. For $C\in\mathbb{R}^{M\times M}$, let $\mathcal{G}(C)$ be the graph with adjacency matrix $A_{ij}=0$, if $C_{ij}=0$ and $A_{ij}=1$, else. Then, the matrix $C$ is irreducible, if and only if $\mathcal{G}(C)$ is strongly connected (see ref.~\cite{Mey2000}, p.~671). For non-negative irreducible matrices $C$, the Perron vector $\mathbf{x}>\mathbf{0}$ is defined by $C\mathbf{x}=\rho(C)\mathbf{x}$ with $\|\mathbf{x}\|_1=1$. The Perron-Frobenius theorem for non-negative irreducible matrices states, that $\mathbf{x}$ exists and is the only 
non-negative eigenvector, except for multiples of $\mathbf{x}$ (\cite{Mey2000}, p.~673). Further, $\rho(C)$ is a simple eigenvalue.

\begin{lemma}\label{lemma:rho_eq_1}
	The Hashimoto matrix $B\geq 0$ be irreducible. Let $\mathbf{p}(\lambda):[0,1]\rightarrow(0,1]^M$ be a continuous parametrization of the occupation probabilities. If a continuous phase transition occurs at $\lambda_c\in[0,1)$, then
	\begin{equation*}
		\rho(\lambda_c) = 1 \, .
	\end{equation*}
\end{lemma}

\textit{Proof.} As in lemma \ref{lemma:rho_geq_1} we may apply the first order expansion for $\mathbf{Q}=\mathbf{1}-\boldsymbol{\epsilon}$
	\begin{equation*}
		\boldsymbol{\epsilon} = \mathbf{F}'(\mathbf{1})\cdot \boldsymbol{\epsilon}+o(\boldsymbol{\epsilon}) \, .
	\end{equation*}	 
Thus, for $\lambda\searrow\lambda_c$, $\boldsymbol{\epsilon}$ converges to a non-negative eigenvector of $\mathbf{F}'(\mathbf{1},\lambda_c)$ with eigenvalue 1. $\mathbf{F}'(\mathbf{1};\lambda_c)\geq 0$ is irreducible, hence, $\boldsymbol{\epsilon}$ converges to a multiple of the Perron vector with eigenvalue $\rho(\lambda_c)=1$. $\square$

Above the percolation threshold, the fixed-point equation possesses a non-trivial solution. For irreducible Hashimoto matrices, this can be shown using the Brouwer fixed-point theorem, which states that any continuous function $f:D\rightarrow D$ on a compact convex subset $D\neq\emptyset$ of 
a finite-dimensional normed vector space has a fixed point (see e.g.~ref.~\cite{Wer2018}, p. 194). 

\begin{theorem}\label{thm:existence}
	The Hashimoto matrix $B$ be irreducible, $\mathbf{p}\in(0,1]^M$ and $\rho>1$. Then, there exists a non-trivial solution $\mathbf{Q}=\mathbf{F}(\mathbf{Q})\in[0,1]^M\setminus\mathbf{1}$.
\end{theorem}
\textit{Proof.} $\mathbf{F}:[0,1]^M\rightarrow [0,1]^M$ is a continuous function on a finite-dimensional normed vector space. Using the Brouwer fixed-point theorem, it is sufficient to find a compact convex subset $D_\delta\subseteq [0,1]^M\setminus \mathbf{1}$, such that $\mathbf{F}(D_\delta)\subseteq D_\delta$. Let $\lambda_i$ be the eigenvalues and $\lambda_1=\rho$. Consider the $M-1$ dimensional affine subspace
\begin{equation*}
U_\delta:=\mathbf{1}-\delta\mathbf{x}+U \qquad U:=\bigoplus_{i\geq 2}V[\lambda_i] \, ,
\end{equation*} 
where $V[\lambda_i]$ are the generalized eigenspaces and $\mathbf{x}>\mathbf{0}$ is the Perron vector for $\mathbf{F}'(\mathbf{1})$. Now we cut off the edge at $\mathbf{1}$ from the domain using the cut surface $U_\delta\cap[0,1]^M$ to obtain the compact convex subset $D_\delta\subseteq [0,1]^M\setminus{\mathbf{1}}$. $\mathbf{F}$ is monotonic, thus, it is sufficient to show that there exists a $\delta>0$, such that $\mathbf{F}$ maps the cut surface to $D_\delta$.
Let $\mathbf{v}\in U_\delta\cap [0,1]^M$ arbitrary, where $\mathbf{v}=:\mathbf{1}-\delta\cdot\mathbf{x}+\mathbf{u}$ with $\mathbf{u}\in U$. Then,
\begin{align*}
		\mathbf{F}(\mathbf{v}) &= \mathbf{1}+\mathbf{F}'(\mathbf{1})\cdot (\mathbf{v}-\mathbf{1})+o(\|\mathbf{v}-\mathbf{1}\|) \\
		&= \mathbf{1}+\mathbf{F}'(\mathbf{1})\cdot (\mathbf{u}-\delta\cdot\mathbf{x})+ o(\delta) \\
		\Leftrightarrow \; \mathbf{F}(\mathbf{v})-\mathbf{F}'(\mathbf{1})\cdot\mathbf{u} &= \mathbf{1}-\delta\cdot\rho\cdot\mathbf{x} + o(\delta) \, .
\end{align*}
Hence, for $\rho>1$ there exists a $\delta>0$, such that
\begin{equation*}
		\mathbf{F}(\mathbf{v})-\mathbf{F}'(\mathbf{1})\cdot\mathbf{u} \quad \in 
D_\delta \, .
\end{equation*}
Since $\mathbf{F}'(\mathbf{1})\cdot\mathbf{u}\in U$ and $\mathbf{F}(\mathbf{v})\in [0,1]^M$, we find
\begin{equation*}
		\mathbf{F}(\mathbf{v})\in D_\delta \, ,
\end{equation*}
which concludes the proof. $\square$

\section{Numerical solutions}\label{sec:numerical_solutions}

\begin{figure}
\includegraphics[width=0.48\textwidth]{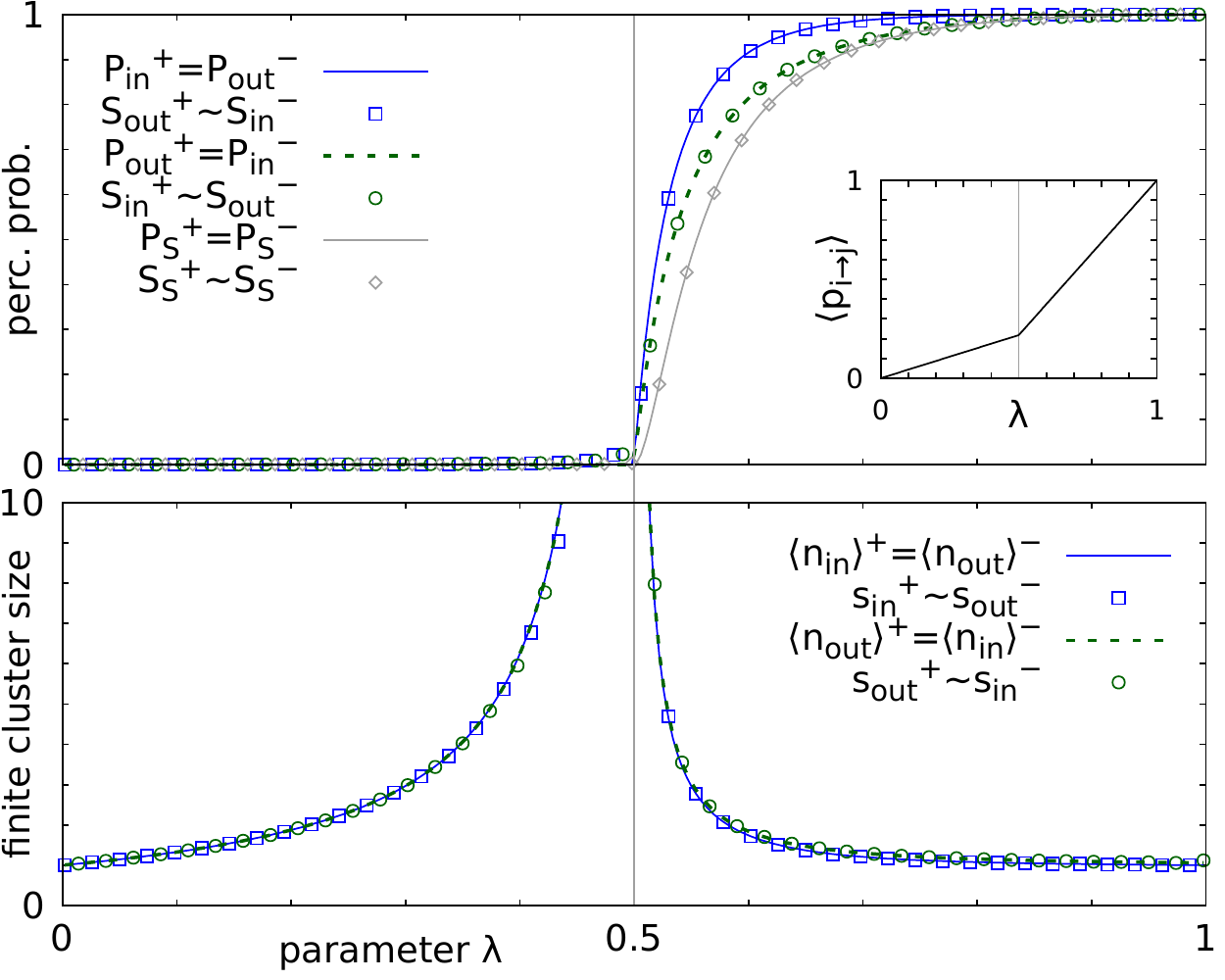}
\caption{Uniform distributed graph with $10^4$ nodes, 60,068 directed edges and degree distribution $f(z)=1/9$ for $z=2,\dots,10$. Theoretical 
results (lines) and simulations (symbols) for the percolation probabilities (top) and the average finite cluster sizes for a randomly chosen node (bottom) for the  parametrization $\mathbf{p}^+$ (labeled with $+$), which yields complementary results with respect to $\mathbf{p}^-$ (labeled with $-$). The vertical line shows the theoretical percolation threshold at 
$\lambda_c=0.5$, see appendix \ref{sec:a_priori_threshold}. }
\label{fig:uniform}
\end{figure}

\begin{figure}
\includegraphics[width=0.48\textwidth]{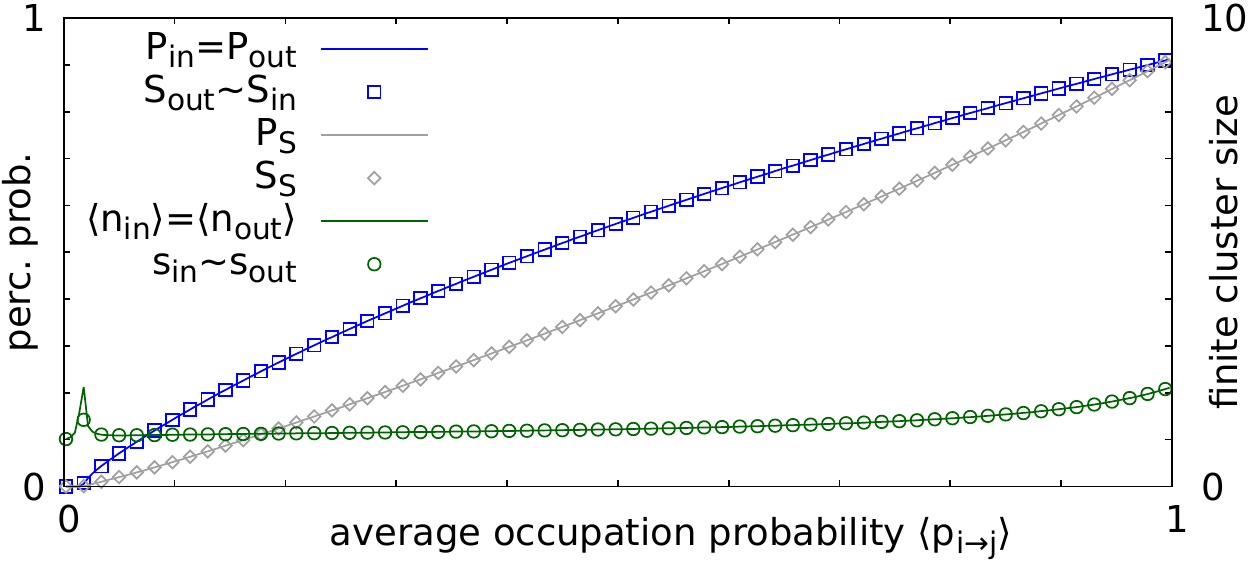}
\includegraphics[width=0.48\textwidth]{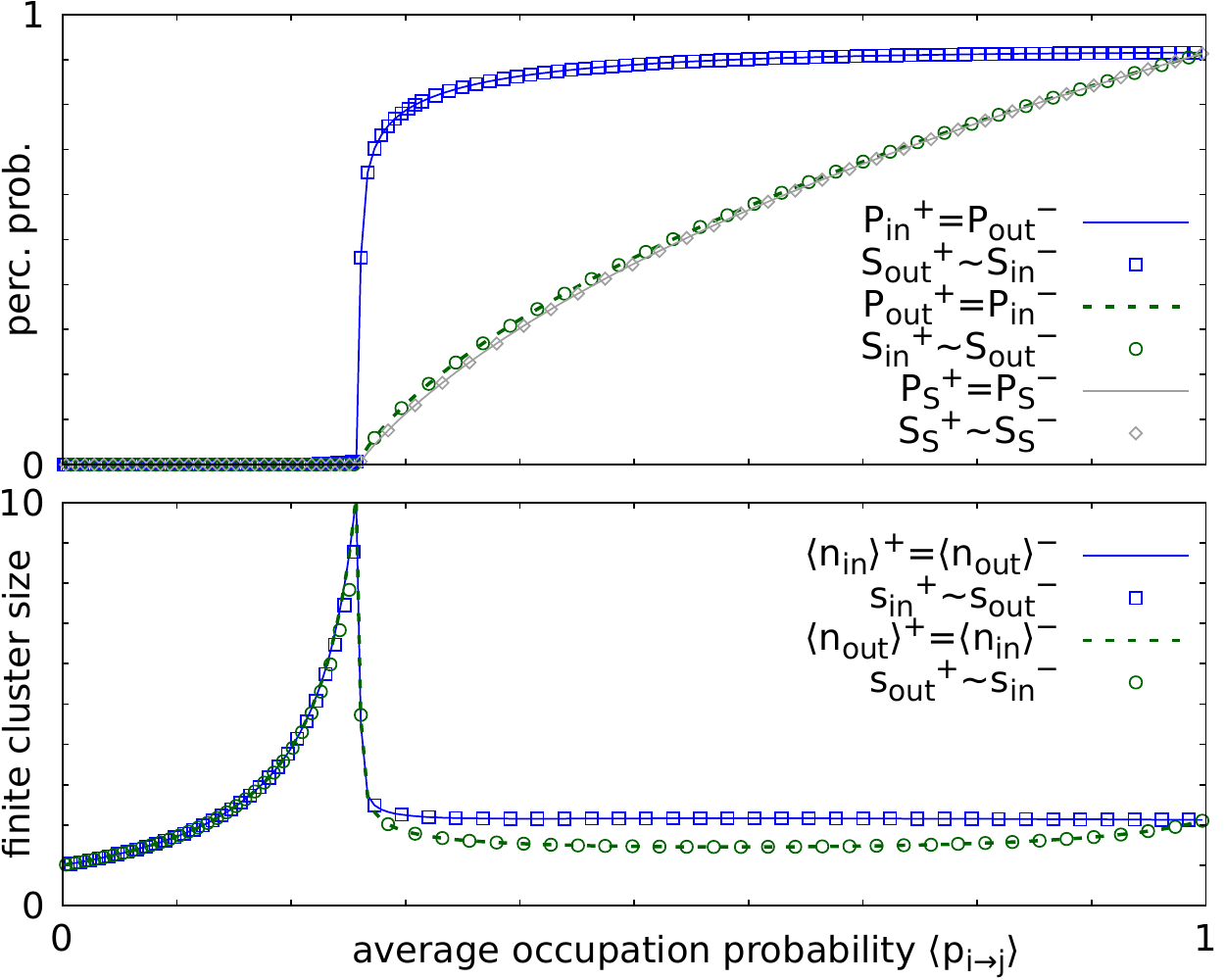}
\caption{Power-law distributed graph with degree distribution $f(z)\propto z^{-2}$ for $z=1,\dots,1000$. Shows the percolation probabilities and 
finite cluster sizes for $p_{i\rightarrow j}\equiv p$ with $10^4$ nodes and 43078 directed edges (top) in comparison to the results for $\mathbf{p}^+$ with $10^5$ nodes and 471234 edges (mid, bottom), which yields complementary results with respect to $\mathbf{p}^-$. }
\label{fig:power_law}
\end{figure}

The directed network obtained by removing all vacant edges can be represented by the bow-tie diagram \cite{Bro2000}, which is widely used to describe the structure of directed networks \cite{New2001,Tim2017,Dor2001,Sch2002,Azi2014}. The giant strongly connected component (GSCC) is defined by the largest strongly connected component. The giant in-component (GIN) is the set of nodes for which a path to GSCC exists and the giant out-component (GOUT) is the set of nodes which can be reached from GSCC, where GIN $\cap$ GOUT $=$ GSCC. The relative sizes of the giant components GSCC, GIN and GOUT are denoted by $S_S$, $S_{\text{in}}$ and $S_{\text{out}}$, respectively. The rest of the network consists of tendrils and disconnected components. 

We require that the sizes of the tendrils and disconnected components are 
small compared to the size of GSCC. Then, for locally tree-like networks, 
the percolation probabilities from equs.~(\ref{equ:P_out}),(\ref{equ:P_in}) and (\ref{equ:P_S}) converge to the relative sizes of the giant components in the large $N$ limit
\begin{equation}\label{equ:asymp_1}
	S_{\text{in}}\sim P_{\text{out}} \quad S_{\text{out}}\sim P_{\text{in}} \quad S_S \sim P_S \, .
\end{equation}
Further, let $s_{\text{out}}$ be the size of clusters of outgoing edges averaged over all nodes which are not part of  GIN and let $s_{\text{in}}$
be the size of clusters of incoming edges averaged over all nodes which are not part of GOUT. Then, 
\begin{equation}\label{equ:asymp_2}
	 s_{\text{out}} \sim  \langle n_{\text{out}}\rangle\quad 
	s_{\text{in}} \sim  \langle n_{\text{in}}\rangle \, ,
\end{equation}
where $\langle n_{\text{out}}\rangle,\langle n_{\text{in}}\rangle$ are the average finite cluster sizes from equs.~(\ref{equ:acs_out_1}) and (\ref{equ:acs_in_1}).

\begin{figure}
\includegraphics[width=0.48\textwidth]{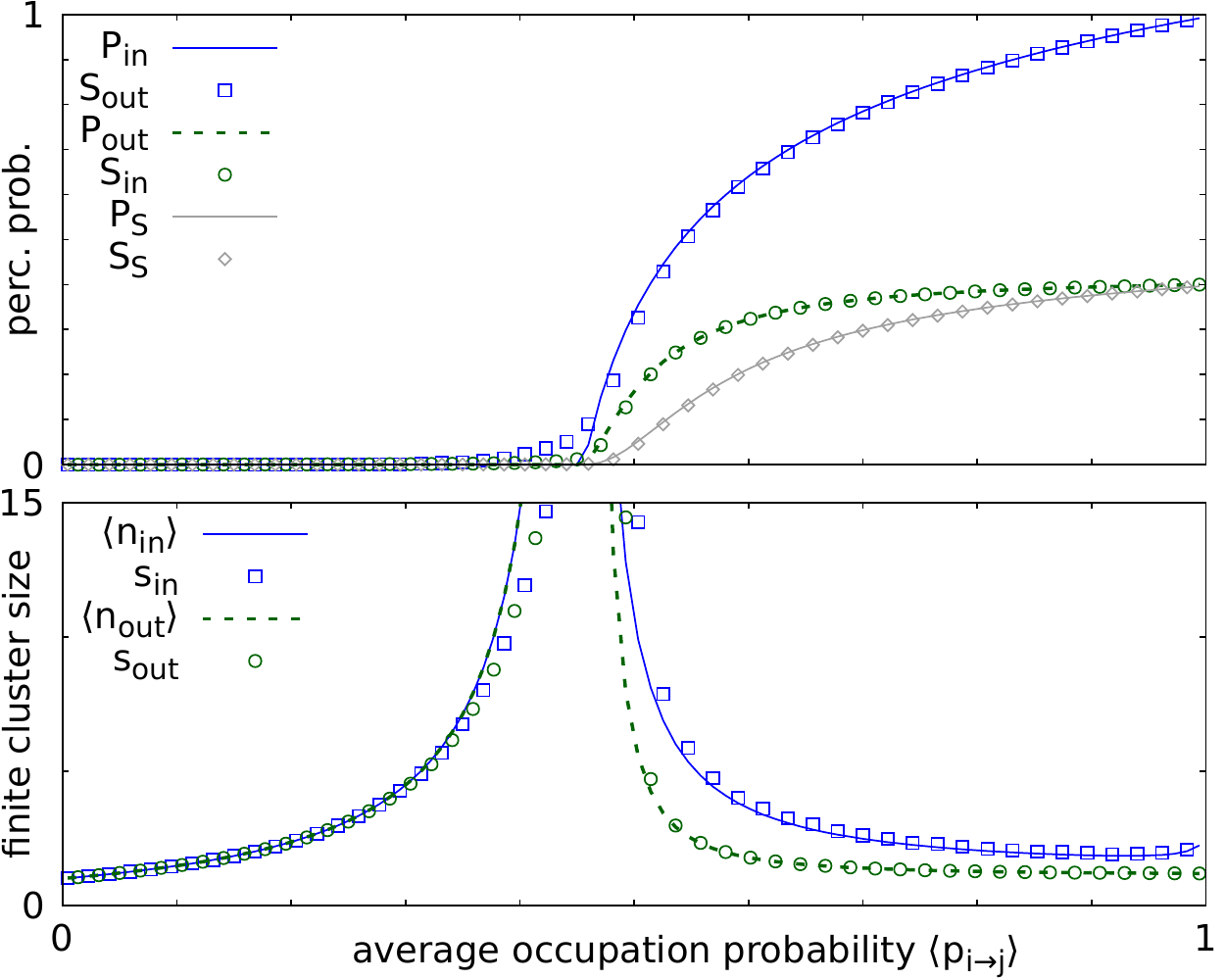}
\caption{Percolation probabilities and finite cluster sizes for the Gnutella peer to peer network \cite{Sta2014} with 10876 nodes and 39994 edges for $\mathbf{p}^-$.}
\label{fig:gnutella}
\end{figure}

\begin{figure}
\includegraphics[width=0.48\textwidth]{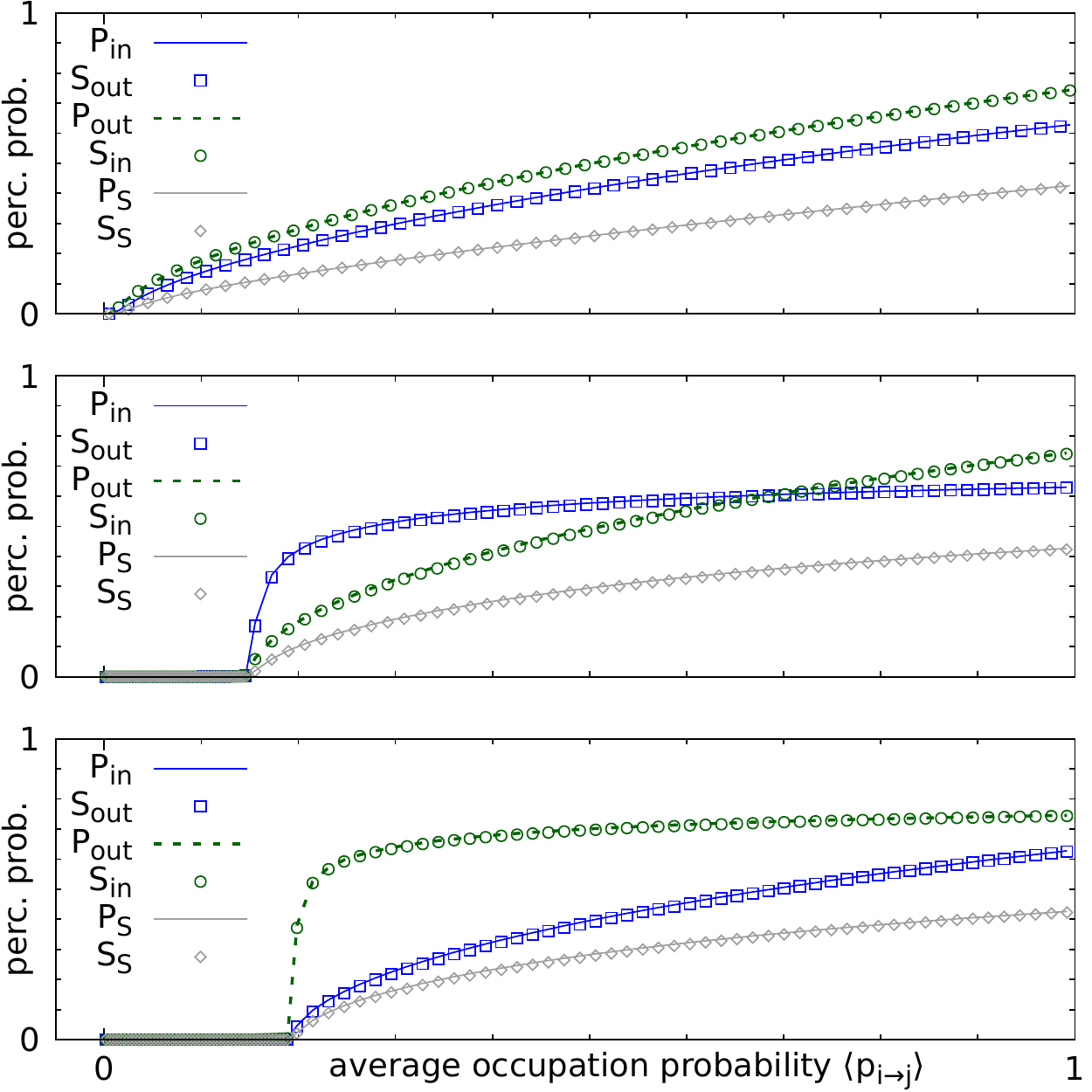}
\caption{Theoretical percolation probabilities (lines) and relative sizes 
of the giant components averaged over 1000 runs (symbols) for the Epinion 
trust network for consumer reviews \cite{Sta2014} with 75879 nodes and 508837 edges for $p_{i\rightarrow j}\equiv p$ (top), $\mathbf{p}^+$ (mid) and $\mathbf{p}^-$ (bottom).}
\label{fig:epinion}
\end{figure}

\begin{figure}
\includegraphics[width=0.48\textwidth]{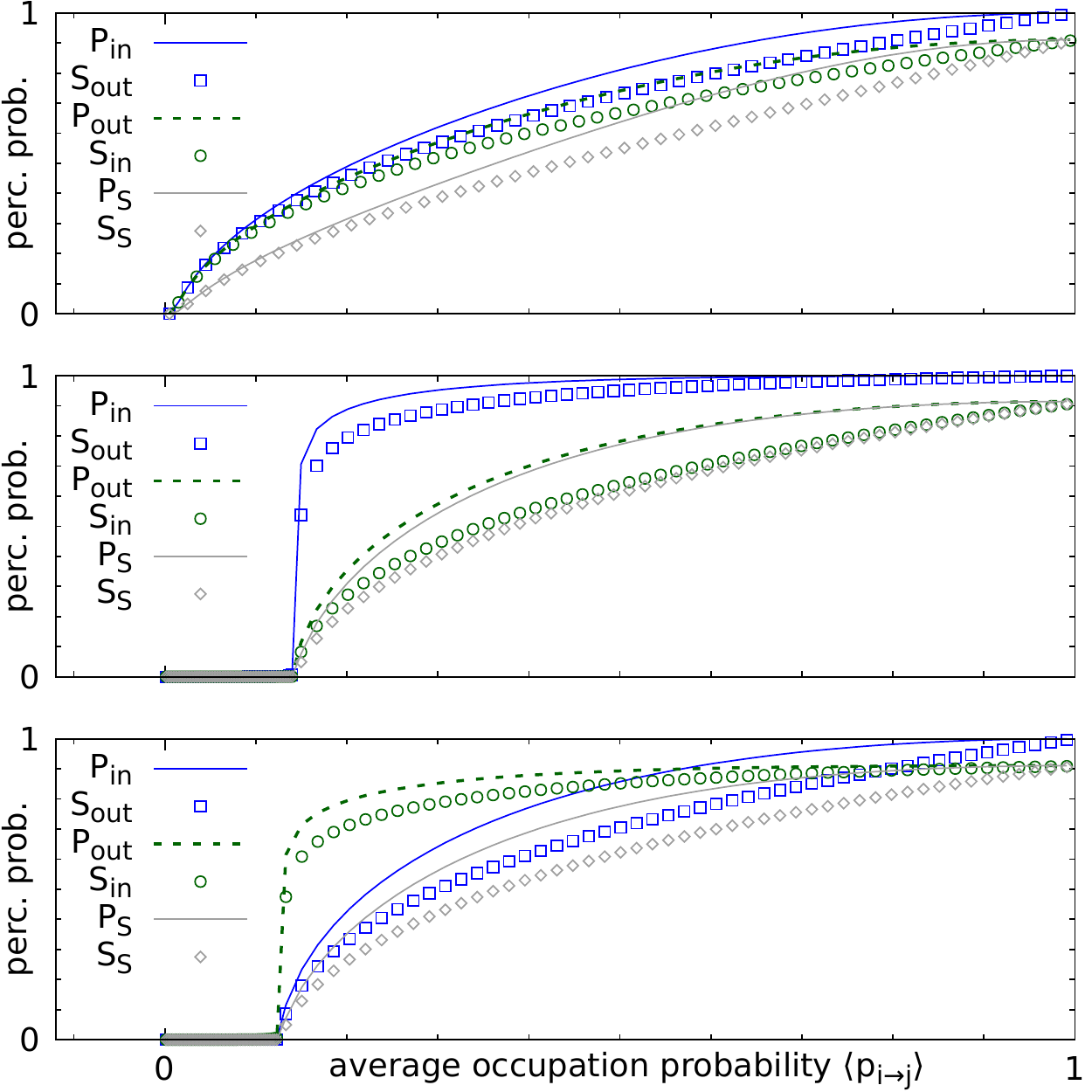}
\caption{Theoretical percolation probabilities (lines) and relative sizes 
of the giant components averaged over 1000 runs (symbols) for the Slashdot network for authors of technology related news consisting of friends and foes \cite{Sta2014} with 77360 nodes and 905468 edges for $p_{i\rightarrow j}\equiv p$ (top), $\mathbf{p}^+$ (mid) and $\mathbf{p}^-$ (bottom).}
\label{fig:slashdot}
\end{figure}

In the following, we investigate the solutions for two non-symmetric parametrizations $\mathbf{p}^+(\lambda),\mathbf{p}^-(\lambda)$ with $\mathbf{p}^\pm(0)=\mathbf{0}$, $\mathbf{p}^\pm(1)=\mathbf{1}$ and linear components, except at $\lambda=0.5$, where 
\begin{align}
	p^+_{i\rightarrow j}(0.5) := |\mathcal{N}^+(j)\setminus i|^{-1} \label{equ:def_p_plus}\\
	p^-_{i\leftarrow j}(0.5) := |\mathcal{N}^-(j)\setminus i|^{-1} \label{equ:def_p_minus}\, .
\end{align}
If $\mathcal{N}^\pm(j)\setminus i = \emptyset$, the corresponding occupation probability at $\lambda=0.5$ is set to one. Figs.~(\ref{fig:uniform}) and (\ref{fig:power_law}) show the solutions for two undirected random graphs, where each edge is decomposed into two anti-parallel edges and 
figs.~(\ref{fig:gnutella})-(\ref{fig:slashdot}) show the solutions for real directed networks from the Stanford collection (SNAP) \cite{Sta2014}. Simulations were averaged over 1000 realizations.

The numerical simulations in figs.~(\ref{fig:uniform}) and (\ref{fig:power_law}) for (locally treelike) undirected random graphs coincide perfectly with the theoretical predictions for the percolation probabilities and average finite cluster sizes, which confirms equs.~(\ref{equ:asymp_1}) and (\ref{equ:asymp_2}). Further, fig.~(\ref{fig:uniform}) shows a uniform distributed graph for which $\rho(0.5)=1$ for both parametrizations, see appendix \ref{sec:a_priori_threshold}. Indeed, the percolation threshold occurs exactly at $\lambda_c=0.5$ in agreement with equs.~(\ref{equ:p_c_in}) and (\ref{equ:p_c_out}). Near the percolation threshold, finite-size effects occur. Here, GSCC is the largest strongly connected component, thus, the relative sizes of the giant components beyond the percolation threshold will only vanish in the large $N$ limit. For random graphs, the finite size yields a chance to encounter small loops, which can be seen in the average cluster size beyond the percolation threshold in figs.~(\ref{fig:uniform}) and (\ref{fig:power_law}), where the theoretical results are an upper bound in agreement to equs.~(\ref{equ:acs_upper_bound_out}) and (\ref{equ:acs_upper_bound_in}), which also hold for real networks, see fig.~(\ref{fig:gnutella}). On large networks, the finite-size effects become negligible, however, the occurrence of loops decreases the percolation probabilities in accordance to equs.~(\ref{equ:upper_bound_out}) and (\ref{equ:upper_bound_in}), see figs.~(\ref{fig:epinion}) and (\ref{fig:slashdot}) for the Epinion and Slashdot network. In contrast to Epinion, the Slashdot network shows large deviations, which is explained by 
a higher average degree resulting in a larger GSCC and significant node overcount due to closed loops. 

Interestingly, the parametrizations $\mathbf{p}^\pm$ significantly delay the formation of giant clusters in comparison to the standard case $p_{i\rightarrow j}\equiv p$. For $\mathbf{p}^+$, the occupation probabilities are anti-correlated with the number of outgoing edges of the end node, which creates a bottleneck for $P_{\text{in}}$. Similar, for $\mathbf{p}^-$, the occupation probabilities are anti-correlated with the number of incoming edges of the starting node, which creates a bottleneck for $P_{\text{out}}$. At the percolation threshold, the bottleneck is overloaded, which 
may induce an abrupt increase of the respective percolation probabilities, see figs.~(\ref{fig:power_law}),(\ref{fig:slashdot}). 

\section{SIR epidemic model}\label{sec:SIR}

Within the scope of the SIR model, each node represents an individual which is either susceptible (S), infected (I) or recovered (R). Each edge $i\rightarrow 
j$ represents a contact through which a transmission might occur. The transmission probability $p_{i\rightarrow j}(\tau_i)$ is the probability that an infected node $i$ transmits the disease to node $j$, if node $j$ is not infected by another neighbor, where $\tau_i$ is the time span for which node $i$ is infectious. At time $\tau_i$ after infection, node $i$ will recover and gain immunity. The recovery times $\tau_i$ are drawn independently from arbitrary distributions $r_i(\tau_i)$. For fixed recovery times, the transmission of the disease along directed edges is assumed to be independent.

\subsection{Late-time behavior}\label{sec:late_time}

First, consider fixed recovery times $\boldsymbol{\tau}$. Then, the contagions are independent and the state of the network at the end of an epidemic can be interpreted as a percolation problem, where the occupation probabilities equal the transmission probabilities $\mathbf{p}(\boldsymbol{\tau})$. Clusters of outgoing edges represent clusters of infected individuals for a singly infected node. Thus, the \textit{major outbreak probability} for node $i$ equals the probability $P_{\text{out}}(i)$ from equ.~(\ref{equ:P_out_i}) that node $i$ is part of a giant cluster of outgoing edges and the major outbreak probability for a randomly chosen node is given by equ.~(\ref{equ:P_out}) for the percolation probability $P_{\text{out}}$. 

Similarly, a node $i$ will contract the disease, if an initially infected 
node is part of the cluster of incoming edges of node $i$. Hence, the probability that node $i$ will be infected during the epidemic, if $(\mathbf{1}-\mathbf{q})\in [0,1)^N$ are the probabilities for each node to be initially infected, is given by the probability $1-Q_{0i}(\mathbf{q})$ that node $i$ is not part of a vacant cluster of incoming edges, where $Q_{0i}$ is given by equs.~(\ref{equ:mp_in_1}) and (\ref{equ:mp_in_2}). Therefore, the total \textit{fraction of infected individuals} is 
\begin{equation*}
	\frac{1}{N}\sum^N_{i=1} 1-Q_{0i}(\mathbf{q}) \, .
\end{equation*}
For a small fraction of initially infected individuals (${\|\mathbf{1}-\mathbf{q}\|_\infty \ll 1}$), the probability that node $i$ will be infected equals the probability $P_{\text{in}}(i)$ from equ.~(\ref{equ:P_in_i}) that node $i$ will be part of a giant cluster 
of incoming edges. Thus, the total fraction of infected individuals for a 
small fraction of initially infected nodes is given by equ.~(\ref{equ:P_in}) for the percolation probability $P_{\text{in}}$. Beyond the percolation threshold, the average outbreak size for a randomly chosen node is given by $\langle n_{\text{out}}\rangle$ from equ.~(\ref{equ:acs_out_4}).

Now, consider the general case where the recovery times $\boldsymbol{\tau}$ are drawn from the distribution $f(\boldsymbol{\tau})=\prod_i r_i(\tau_i)$. Then, the \textit{conditional} PGFs from equs.~(\ref{equ:def_G0}),(\ref{equ:def_G}),(\ref{equ:def_F0}) and (\ref{equ:def_F}) depend on the 
random variable $\boldsymbol{\tau}$ and must be replaced by their average
\begin{align*}
	\overline{F}_{0i}(\mathbf{x})   &:= F_{0i}(\mathbf{x};\overline{\mathbf{p}}) \\
	\overline{F}_{i\rightarrow j}(\mathbf{x}) &:= F_{i\rightarrow j}(\mathbf{x};\overline{\mathbf{p}})
\end{align*}
\begin{align*}
\overline{G}_{0i}(\mathbf{x}) 
&:=\int^\infty_0 r_i(\tau_i) \cdot G_{0i}(\mathbf{x}|\tau_i) \, d\tau_i 
\\
	\overline{G}_{i\rightarrow j}(\mathbf{x})  
	&:= \int^\infty_0 r_j(\tau_j)\cdot G_{i\rightarrow j}(\mathbf{x}|\tau_j) \, d\tau_j \, ,
\end{align*}
With these definitions the interpretation in this section remains valid and one obtains the exact results on large locally treelike networks. This 
can be shown either by repeating the derivation in sec.~\ref{sec:message_passing} using the new PGFs or by averaging over the recovery times to obtain the exact solutions on the infinite trees $\mathcal{G}(i)$ and $\mathcal{F}(i)$, where the recovery times are drawn independently for each node from the infinite trees
\begin{align}
	\overline{Q}_{0i}(\mathbf{x}) &= Q_{0i}(\mathbf{x};\overline{\mathbf{p}}) \label{equ:sir_mp1}\\
	\overline{H}_{0i}(\mathbf{x}) &= x_i\cdot\overline{G}_{0i}(\overline{\mathbf{H}}(\mathbf{x})) \label{equ:sir_mp2}\\
	\overline{H}_{i\rightarrow j}(\mathbf{x}) &= x_j\cdot \overline{G}_{i\rightarrow j}(\overline{\mathbf{H}}(\mathbf{x})) \label{equ:sir_mp3}\, .
\end{align}
Thus, non-degenerate recovery times $\boldsymbol{\tau}$ result in a lack of symmetry due to statistically dependent transmission probabilities (which was previously shown for the SIR model on random graphs \cite{Ken2007}).
The major outbreak probability $\overline{P}_{\text{out}}$ for a randomly 
chosen node and the fraction of infected individuals $\overline{P}_{\text{in}}$ for a small fraction of initially infected nodes are given by 
\begin{align}
	\overline{P}_{\text{out}} &= \frac{1}{N}\sum^N_{i=1} 1-\overline{H}_{0i}(\mathbf{1}) \\
	\overline{P}_{\text{in}} &= \frac{1}{N}\sum^N_{i=1} 1-Q_{0i}(\mathbf{1};\overline{\mathbf{p}}) \, .
\end{align}
Similar to equs.~(\ref{equ:p_c_in}) and (\ref{equ:p_c_out}), the epidemic 
threshold at which the fraction of infected individuals as well as the major outbreak probability become positive satisfies $\rho(\mathbf{\overline{F}}'(\mathbf{1}))=\rho(\mathbf{\overline{G}}'(\mathbf{1}))=1$ with
\begin{align*}	
\mathbf{\overline{F}}'(\mathbf{1})=B^T\cdot\diag(\overline{\mathbf{p}}) 
\\
\mathbf{\overline{G}}'(\mathbf{1})=B\cdot\diag(\overline{\mathbf{p}}) \, .
\end{align*}
Since $\overline{G}_{0i}'(\mathbf{1})=G_{0i}'(\mathbf{1};\overline{\mathbf{p}})$ and $\overline{\mathbf{G}}'(\mathbf{1})=\mathbf{G}'(\mathbf{1};\overline{\mathbf{p}})$, the average outbreak size for a randomly chosen node beyond the percolation threshold is given by
\begin{equation*}
	\overline{\langle n_{\text{out}}\rangle} = \frac{1}{N}\sum^N_{i=1}H_{0i}'(\mathbf{1};\overline{\mathbf{p}}) \, ,
\end{equation*}
see equs.~(\ref{equ:acs_out_4})-(\ref{equ:acs_out_6}).

On large networks with loops, the solutions we have given remain a lower bound for the epidemic threshold as well as an upper bound for the average outbreak size beyond the epidemic threshold, the major outbreak probability and the fraction of infected individuals. Similar to ref.~\cite{Ken2007}, by recursively applying Jensen's inequality to equs.~(\ref{equ:sir_mp2}) and (\ref{equ:sir_mp3}), we find
\begin{align*}
	\overline{H}_{0i}(\mathbf{x}) & \geq H_{0i}(\mathbf{x};\overline{\mathbf{p}}) \, .
\end{align*}
where we assume the convergence for the initial value $\overline{\mathbf{H}}(\mathbf{x})$ to the fixed-point $\mathbf{H}(\mathbf{x};\overline{\mathbf{p}})$. Hence, using the PGFs from equs.~(\ref{equ:def_G0}),(\ref{equ:def_G}),(\ref{equ:def_F0}) and (\ref{equ:def_F}) for bond percolation, the occupation probabilities can be chosen to be $\mathbf{p}=\overline{\mathbf{p}}$, which yields the correct results for the fraction of infected individuals, the average outbreak sizes beyond the epidemic threshold as well as the epidemic threshold but an upper bound for the major outbreak probability on large locally treelike networks in agreement with the results from ref.~\cite{Ken2007}. 

\subsection{Suppression of large outbreaks}\label{sec:suppression_large_outbreaks}
The occupation probabilities be defined by $\mathbf{p}:=\overline{\mathbf{p}}$ as described in the previous section. For any induced matrix norm, one obtains two criteria, which  prohibit large outbreaks and guarantee 
vanishing percolation probabilities 
\begin{align*}
	\|\mathbf{G}'(\mathbf{1})\| &< 1 \\
	\|\mathbf{F}'(\mathbf{1})\| &< 1 \, .
\end{align*}
According to equs.~(\ref{equ:p_c_in}),(\ref{equ:p_c_out}) and (\ref{equ:rho}), both criteria yield $\rho<1$, hence, $P_{\text{in}}=P_{\text{out}}=0$. For the row-sum and column-sum norm, we find
\begin{align*}
	\forall_{i\leftarrow j}\; p_{i\leftarrow j} < \frac{1}{|\mathcal{N}^-(j)\setminus i|} \quad &\Rightarrow\quad \|\mathbf{G}'(\mathbf{1})\|_1 < 1 \\
	\forall_{i\rightarrow j}\; p_{i\rightarrow j} < \frac{1}{|\mathcal{N}^+(j)\setminus i|} \quad &\Rightarrow\quad \|\mathbf{F}'(\mathbf{1})\|_1 < 1 
\\
	\forall_{i\rightarrow j}\; \sum_{k\in \mathcal{N}^+(j)\setminus i} p_{j\rightarrow k} < 1 \quad &\Rightarrow\quad \|\mathbf{G}'(\mathbf{1})\|_\infty < 1 \\
\forall_{i\leftarrow j}\;	\sum_{k\in \mathcal{N}^-(j)\setminus i} p_{j\leftarrow k} < 1 \quad &\Rightarrow\quad \|\mathbf{F}'(\mathbf{1})\|_\infty 
< 1 
\end{align*}
For convenience, let each edge possess an anti-parallel edge, such that $\mathcal{N}(i)=\mathcal{N}^\pm(i)$. If one of the following statements holds for each node $j$
\begin{align*}
	&\max_{i\in\mathcal{N}(j)} p_{i\leftarrow j} < \frac{1}{|\mathcal{N}(j)|-1} \\
	&\max_{i\in\mathcal{N}(j)} p_{i\rightarrow j} < \frac{1}{|\mathcal{N}(j)|-1} \\
	&\max_{i\in\mathcal{N}(j)} \sum_{k\in \mathcal{N}(j)\setminus i} p_{j\rightarrow k} < 1 \\
	&\max_{i\in\mathcal{N}(j)} \sum_{k\in \mathcal{N}(j)\setminus i} p_{j\leftarrow k} < 1 \; ,
\end{align*}
then $P_{\text{in}}=P_{\text{out}}=0$. Hence, major outbreaks due to the infection of a single node become impossible. Further, if a small fraction of the network is initially infected, only a small fraction of the population will be infected at the end of the epidemic. Since the message-passing approximation yields an upper bound for the percolation probabilities, these criteria hold for any large network, regardless the existence of many small loops. For $|\mathcal{N}(j)|\geq 2$ the first two criteria are sharp on large locally tree-like networks, except for increments of the transmission probabilities which leave $\rho$ invariant, see appendix \ref{sec:a_priori_threshold}.

\section{Conclusion}\label{sec:conclusion}
In summary, we derived the percolation transition and generalized message-passing equations for the cluster size distribution on weighted, directed networks by extending the generating function formalism in analogy to the theory of random graphs. These equations determine upper bounds for the percolation probabilities (and hence a lower bound for the percolation threshold), which become exact for locally tree-like networks. Numerical simulations on large random graphs with asymmetric occupation probabilities accurately confirm the theoretical predictions for the percolation probability, percolation transition and average cluster size. We demonstrated that the message-passing approximation on real directed networks still is in very good agreement with numerical simulations, if the network is large and sparse. On scale-free and social networks we observed an eminent increase of the percolation threshold, if the occupation probabilities are anti-correlated with the degree of the start and end node, which induces a bottleneck for the size of the giant in- and out-component, respectively. Further we discussed the SIR model on weighted, directed networks and have given a lower bound for the epidemic threshold as well as upper bounds for the average outbreak size, the major outbreak probability and the fraction of infected individuals, and we have proposed strategies to suppress major outbreaks (``vaccination strategies''). The derivation naturally includes modified message-passing equations which remain exact on large locally tree-like networks by taking into account correlations between transmission probabilities due to non-degenerate recovery times. 

\appendix
\section{}\label{sec:a_priori_threshold}

We prove that
\begin{align*}
	\forall_{i\leftarrow j} \quad p_{i\leftarrow j} = |\mathcal{N}^-(j)\setminus i|^{-1} \quad &\Rightarrow\quad \rho = 1 \\
		\forall_{i\rightarrow j} \quad p_{i\rightarrow j} = |\mathcal{N}^+(j)\setminus i|^{-1} \quad &\Rightarrow\quad \rho = 1 \, .
\end{align*}
We consider only the first statement, since the second is derived the same way. 

\textit{Proof.} It is easy to show that 
\begin{equation*}
	\forall_{i\leftarrow j} \quad p_{i\leftarrow j} < |\mathcal{N}^-(j)\setminus i|^{-1} \quad\Rightarrow\quad \rho \leq \|\mathbf{G}'(\mathbf{1})\|_1 < 1 \, .
\end{equation*}
Since the spectral radius is continuous, it is left to prove that
\begin{equation*}
\forall_{i\leftarrow j} \quad p_{i\leftarrow j} > |\mathcal{N}^-(j)\setminus i|^{-1} \quad \Rightarrow\quad \rho \geq 1 \, .
\end{equation*}
Taking the first order expansion, we have
\begin{align*}
	&\mathbf{G}(\mathbf{1}-\delta\mathbf{e}_{k\rightarrow l})=\mathbf{1}-\delta\cdot p_{k\rightarrow l}\sum_{i\rightarrow j} B_{i\rightarrow j, k\rightarrow l}\cdot\mathbf{e}_{i\rightarrow j} + o(\delta) \\
		& \bigg\|\sum_{i\rightarrow j} B_{i\rightarrow j, k\rightarrow l}\cdot\mathbf{e}_{i\rightarrow j}\bigg\|_1 = |\mathcal{N}^-(k)\setminus l| \, .
\end{align*}
Thus,
\begin{align*}
	&\forall_{i\leftarrow j} \, p_{i\leftarrow j} > |\mathcal{N}^-(j)\setminus i|^{-1} \, \Rightarrow\, \exists_{\delta_0>0}\forall_{0<\delta<\delta_0} \mathbf{G}(C_\delta)\subseteq C_\delta \\
	&C_\delta :=\{\mathbf{y}\in [0,1]^M \,|\,\|\mathbf{1}-\mathbf{y}\|_1 \geq \delta \} \, .
\end{align*}
Now, assume $\rho < 1$. Then, using the Perron vector $\mathbf{x}\geq\mathbf{0}$ for $\mathbf{G}'(\mathbf{1})$, we find a contradiction to the previous statement
\begin{align*}
	&\mathbf{G}(\mathbf{1}-\delta\mathbf{x}) = \mathbf{1}-\delta\rho\mathbf{x}+o(\delta) \\
	&\Rightarrow\quad \exists_{\delta_0>0}\forall_{0<\delta<\delta_0} \|\mathbf{1}-\mathbf{G}(\mathbf{1}-\delta\mathbf{x})\|_1 < \delta \\
	&\Rightarrow\quad \exists_{\delta_0>0}\forall_{0<\delta<\delta_0} (\mathbf{1}-\delta\mathbf{x})\in C_\delta \land \mathbf{G}(\mathbf{1}-\delta\mathbf{x}) \notin C_\delta \, . \,\square
\end{align*}

Therefore, if $\mathcal{N}^\pm(j)\setminus i\neq\emptyset$ for all edges $i\rightarrow j$, we have $\rho(0.5)=1$ for the parametrizations $\mathbf{p}^\pm$ from equs.~(\ref{equ:def_p_plus})(\ref{equ:def_p_minus}). Assuming that the spectral radius is strict monotonic near $\lambda=0.5$, equs.~(\ref{equ:p_c_in})(\ref{equ:p_c_out}) predict a phase transition at $\lambda_c=0.5$, see fig.~(\ref{fig:uniform}).

\begin{acknowledgments}
The authors thank Peter Pfaffelhuber for helpful discussions. This work was supported by the state of Baden-Württemberg through bwHPC
and the German Research Foundation (DFG) through grant no INST 39/963-1 FUGG (bwForCluster NEMO). T.S. acknowledges funding by the German Research Foundation in project 404913146.
\end{acknowledgments}


%

\end{document}